\documentclass[twocolumn]{aastex631}

\usepackage{amsmath}

\shorttitle{Nonthermal Acceleration of particles at a Nonrelativistic Quasiparallel Shock}
\shortauthors{Yu et al.}
\graphicspath{{./}{figures/}}

\begin{document}

\title{Nonthermal Acceleration of Electrons, Positrons and Protons at a Nonrelativistic Quasiparallel Collisionless Shock}

\correspondingauthor{Jun Fang}
\email{fangjun@ynu.edu.cn}

\author{Huan Yu}
\affiliation{Department of Physical Science and Technology, Kunming University, Kunming 650214, China}
\author{Qi Xia}
\author[0000-0001-8043-0745]{Jun Fang}
\affiliation{Department of Astronomy, School of Physics and Astronomy, Key Laboratory of Astroparticle Physics of Yunnan Province, Yunnan University, Kunming 650091, People's Republic of China}

\begin{abstract}

Energetic positrons have been observed in the interstellar medium, and high-energy positrons with relativistic energies up to approximately 1 TeV have been detected in Galactic cosmic rays. We conducted a study on the acceleration of particles, specifically positrons, in a nonrelativistic quasiparallel collisionless shock induced by a plasma consisting of protons, electrons, and positrons. The positron-to-proton number density ratio in the plasma is 0.1. We focused on a representative shock with a sonic Mach number of 17.1 and an Alfv\'{e}nic Mach number of 16.8 in the rest frame of the shock. To investigate the acceleration mechanisms of particles including positrons in the shock, we utilized one-dimensional particle-in-cell (PIC) simulations.
It was found that all three species of particles in the shock can be accelerated and exhibit power law spectra. At the shock front, a significant portion of incoming upstream particles are reflected and undergo significant energy increase, and these reflected particles can be efficiently injected into the process of diffusive shock acceleration (DSA). Moveover, the reflected positrons can be further accelerated by an electric field parallel to the magnetic field when they move along the magnetic field upstream of the shock. As a result, positrons can be preferentially accelerated to be injected in the DSA process compared to electrons.
\end{abstract}


\section{Introduction}
\label{sec:Intro}

DSA has been widely acknowledged as an efficient mechanism for accelerating particles to relativistic energies within shock waves. This acceleration process involves particles repeatedly crossing back and forth through the shock wave surface, gaining energy with each passage. For particles, especially electrons, to undergo DSA, they need to be pre-accelerated to acquire enough energy to diffuse across the shock. This pre-acceleration is necessary because the shock front acts as a barrier that particles need to overcome in order to enter the acceleration process \citep{2009A&ARv..17..409T,2023PPCF...65a4002B}..

Kinetic simulations on particle acceleration in nonrelativistic quasi-parallel shocks suggest that protons and electrons can be accelerated to form a nonthermal tail in the energy/momentum spectra. Based on 1D long PIC simulations, power-law spectra with an index of approximately -4 in momentum space are formed for protons and electrons in the downstream region of the shock \citep{2015ApJ...802..115K,2015PhRvL.114h5003P,2019RAA....19..182F}.
Protons can readily enter the DSA regime after several gyrocycles of shock drift acceleration (SDA). On the other hand, electrons require pre-acceleration to increase their energy by several orders of magnitude so that their Larmor radii become comparable to the shock width, enabling them to undergo DSA. The 1D simulations indicate that electrons can get trapped between the shock and the upstream region adjacent to the shock through magnetic mirroring and scattering with the upstream waves. These electrons can undergo pre-acceleration via SDA and scattering processes \citep{2015ApJ...802..115K,2015PhRvL.114h5003P}.
The dynamics of the shock and particle acceleration in quasi-parallel shocks induced by a plasma composed of electrons, protons, and helium ions have been investigated. It has been suggested that all three species of particles can be accelerated and injected into the DSA process \citep{2022MNRAS.512.5418F}.

For quasi-perpendicular shocks, protons are more likely to advect downstream along the magnetic field at the shock, and they cannot undergo enough cycles of SDA to reach sufficient energy for injection into the DSA process \citep{2014ApJ...783...91C}. However, electrons can be reflected through magnetic mirroring at the quasi-perpendicular shocks, and nonresonant waves are excited by these reflected electrons in the upstream. As a result, the electrons can undergo enough cycles of SDA to be injected into the DSA process when trapped between the shock and the upstream waves \citep{2020ApJ...897L..41X}. Kinetic simulations have also shown that electron injection and acceleration in quasi-perpendicular shocks are more efficient than proton \citep{2020ApJ...897L..41X}.

In this paper, we conducted research on the acceleration of particles, specifically positrons, in a non-relativistic quasi-parallel shock using one-dimensional PIC simulations. The setup of the simulation is presented in Section \ref{sec:numerics}. In Section \ref{sec:results}, we analyze the distribution of particles and magnetic fields, as well as the particle spectra obtained from the simulation. Lastly, Section \ref{sec:summary} offers a concise summary and discussion of our findings.

\section{Setup of the shock simulation}
\label{sec:numerics}

To investigate the acceleration processes in a non-relativistic quasi-parallel collisionless shock, we conducted simulations using the PIC code Smilei \citep{2018CoPhC.222..351D} in a 1D3V (one spatial dimension and three momentum dimensions) setting. Initially, a plasma consisting of protons, electrons, and positrons moved in the negative x-direction with a bulk velocity of $V_0=0.2c$ (where $c$ is the speed of light). A shock was set up after the upstream plasma reflected at the leftmost boundary at $x=0$. The shock propagated along the x-direction and was modeled in the downstream frame, where the downstream plasma had a bulk velocity of 0. In the shock rest frame, the upstream plasma flowed towards the shock front with a velocity of $v_{\mathrm{sh}}=V_0 r/(r-1)=0.27c$, corresponding to a compression ratio of $r=4$. Initially, all three species of particles were in thermal equilibrium with a temperature of $T_{\mathrm{up}}=2\times10^{-3}\,m_{\mathrm{e}}c^2/k_{\mathrm{B}}$, where $m_{\mathrm{e}}$ is the electron mass and $k_{\mathrm{B}}$ is the Boltzmann constant. The magnetic field in the initial upstream plasma was set as $\mathbf{B} = B_0(\cos\theta \hat{\bf{x}} + \sin\theta \hat{\bf{y}})$, with $B_0 = 90\,\mu \mathrm{G}$ and $\theta = 30 ^{\circ}$.

The initial number density of upstream protons is $n_{\mathrm{p}} = n_0 \equiv 0.1\,\mathrm{cm}^{-3}$, while the positrons have a density of $n_{e^{+}} = 0.1n_0$. To maintain electrical neutrality, the density of upstream electrons is set to $n_{e} = 1.1n_0$. The average molecular mass of the plasma is $m_0 = 0.46\,m_{\mathrm{p}}$. In the simulation, we use a reduced proton-to-electron mass ratio of $m_{\mathrm p}/m_{e} = 30$, resulting in an upstream sound speed of $v_{\mathrm{s}} \equiv \sqrt{\gamma k_{\mathrm{B}} T_{\mathrm{up}} /m_0} = 1.57\times 10^{-2}\,c$, where $\gamma = 5/3$ is the adiabatic index. The  Alfv\'{e}n  speed is given by $v_{\mathrm{A}} \equiv B/[4\pi(n_e m_e + n_{e^+} m_e + n_{\mathrm{p}}m_{\mathrm{p}})]^{1/2} = 1.6\times 10^{-2}\,c$. In the shock rest frame, the corresponding sonic Mach number and  Alfv\'{e}nic  Mach number are $M_{\mathrm{s}} = 17.1$ and $M_{\mathrm{A}} = 16.8$, respectively. The simulation domain spans a distance of $6.4\times10^4 c/\omega_{\mathrm{pe}}$, with 64 particles per cell per species.  The spatial resolution is $\delta x = 0.05c/\omega_{\mathrm{pe}}$, and the time resolution is $\delta t = 0.02\,\omega_{\mathrm{pe}}^{-1}$, where $\omega_{\mathrm{pe}}^{-1} = \sqrt{m_{\mathrm{e}}/4\pi n_0 e^2}$ represents the time unit ($-e$ denotes the electron charge).

\section{Results}
\label{sec:results}
\begin{figure}
        \centering
        \includegraphics[width=0.5\textwidth]{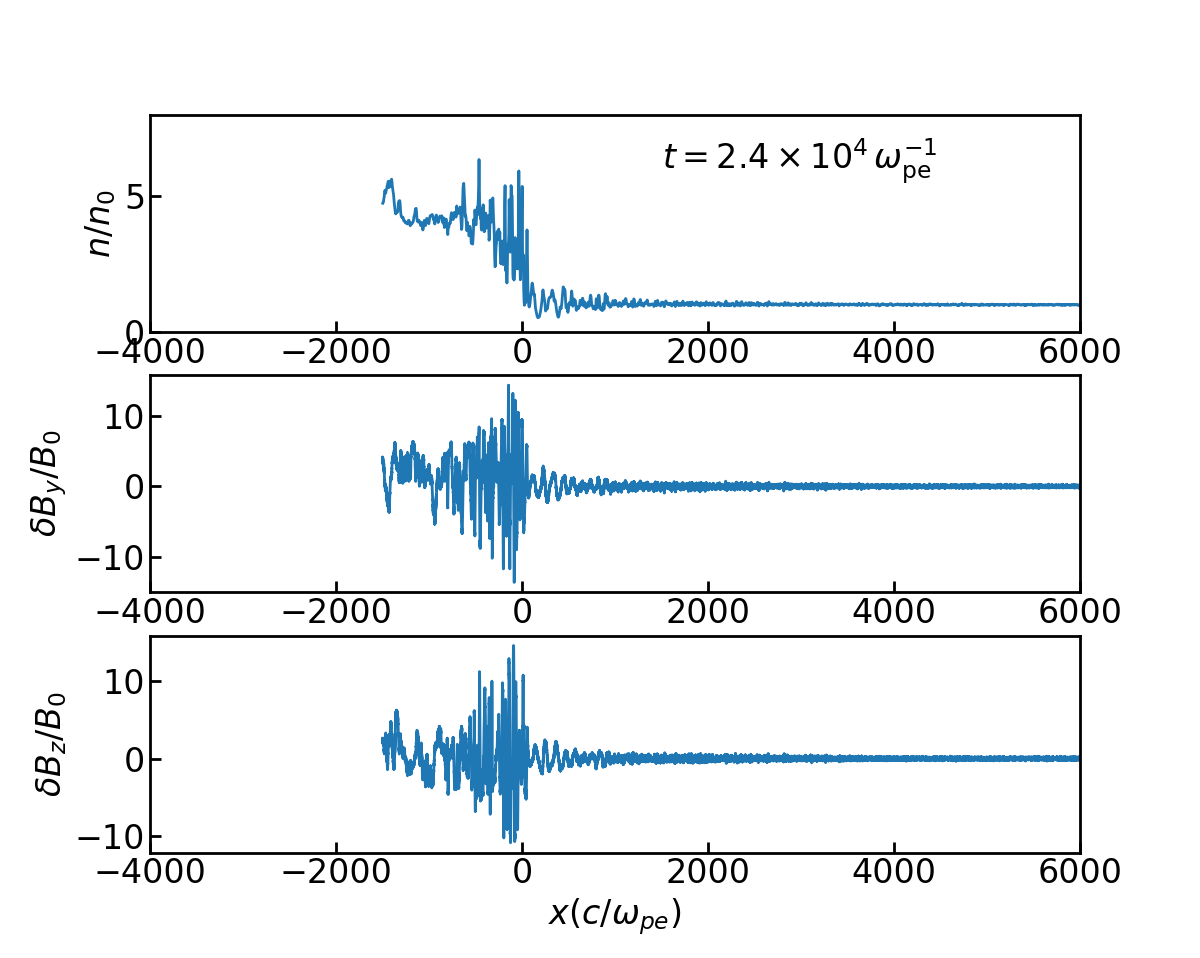}
        \caption{Proton number density normalized by $n_0$ (top panel), the self-generated magnetic field $\delta B_y$ (middle panel) and $\delta B_z$ (bottom panel) for the shock at $t=2.4\times10^4\,\omega_{\mathrm{pe}}^{-1}$. The shock is located at $x_{\mathrm{sh}}=1.5\times 10^2\,c/\omega_{\mathrm{pe}}$, and the $x$ coordinate is shifted by setting  the shock location as the origin. }
        \label{fig:shockstruc24}
\end{figure}

\begin{figure}
        \centering
        \includegraphics[width=0.5\textwidth]{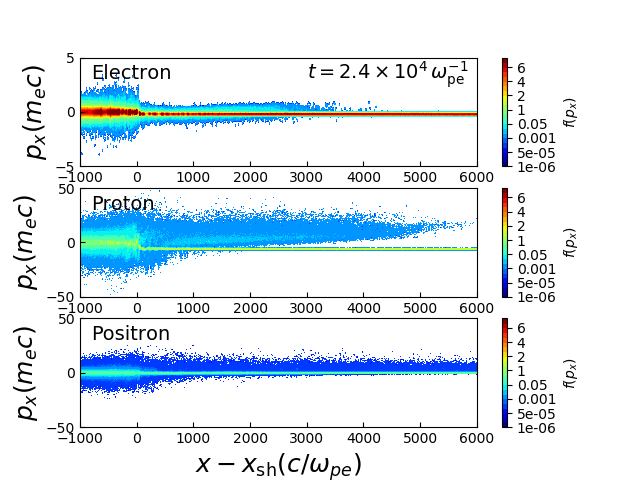}
        \caption{Phase space distributions of the electrons (top panel), the protons (middle panel), and the positrons (bottom panel) for the shock at $t=2.4\times10^4\,\omega_{\mathrm{pe}}^{-1}$ (left panels).}
        \label{fig:phase28}
\end{figure}

\begin{figure*}
        \centering
        \includegraphics[width=1.0\textwidth]{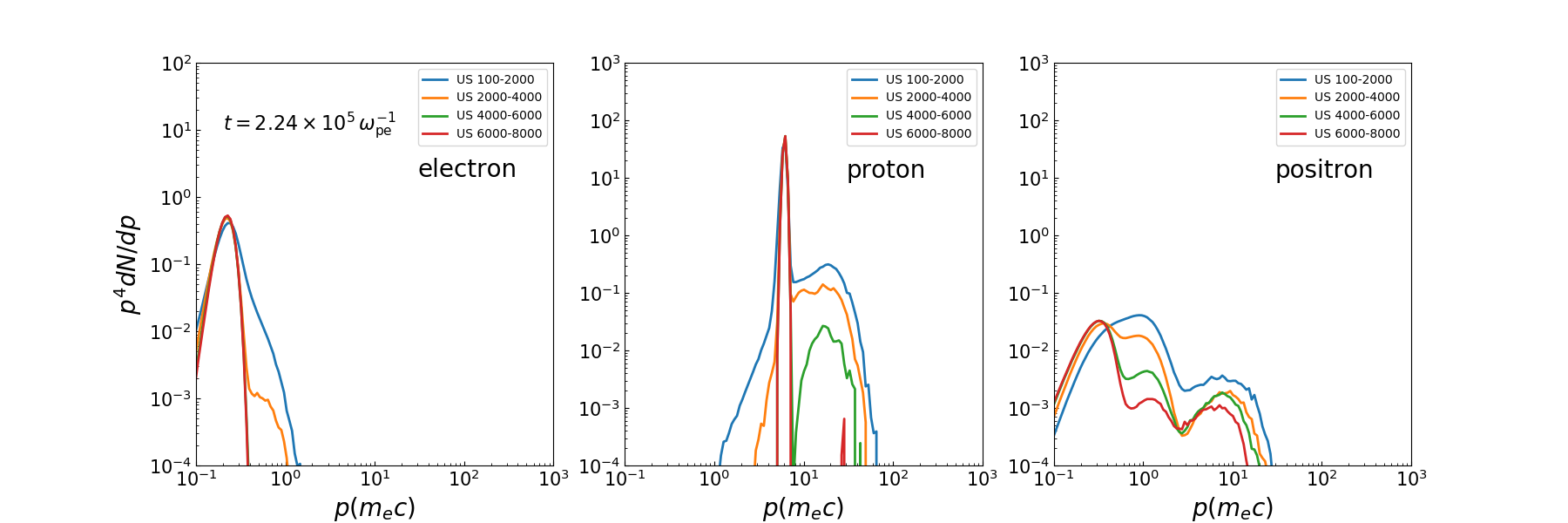}
        \caption{Spectra of electrons (left), protons (middle), and positrons (right), in the region $100-2000\,c/\omega_{\mathrm{pe}}$, $2000-4000\,c/\omega_{\mathrm{pe}}$, $4000-6000\,c/\omega_{\mathrm{pe}}$, $6000-8000\,c/\omega_{\mathrm{pe}}$ before (upstream) the shock at $t=2.4\times10^4\,\omega_{\mathrm{pe}}^{-1}$, respectively. }
        \label{fig:ParticleDis24}
\end{figure*}

\begin{figure*}
        \centering
        \includegraphics[width=0.49\textwidth]{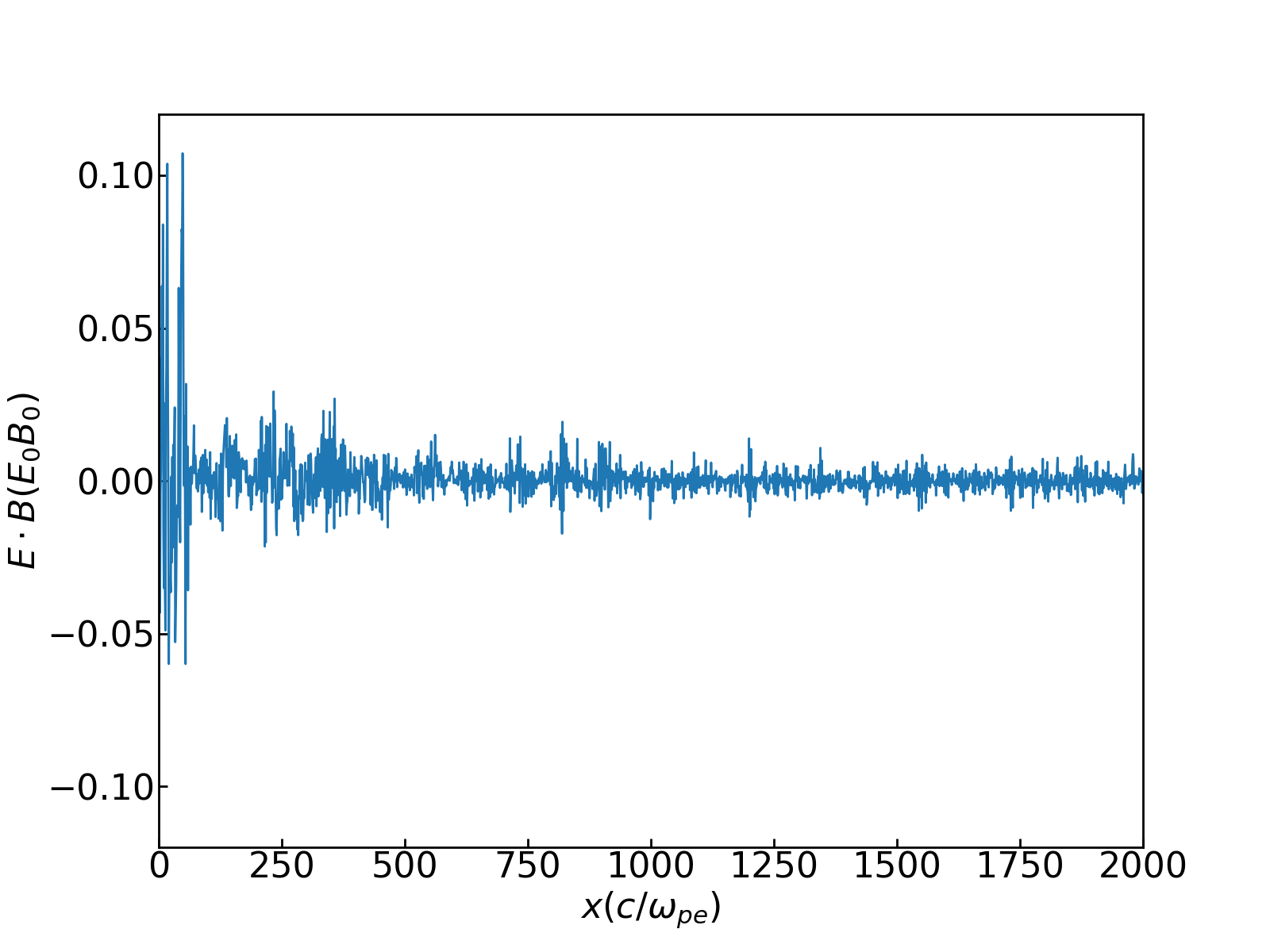}
        \includegraphics[width=0.49\textwidth]{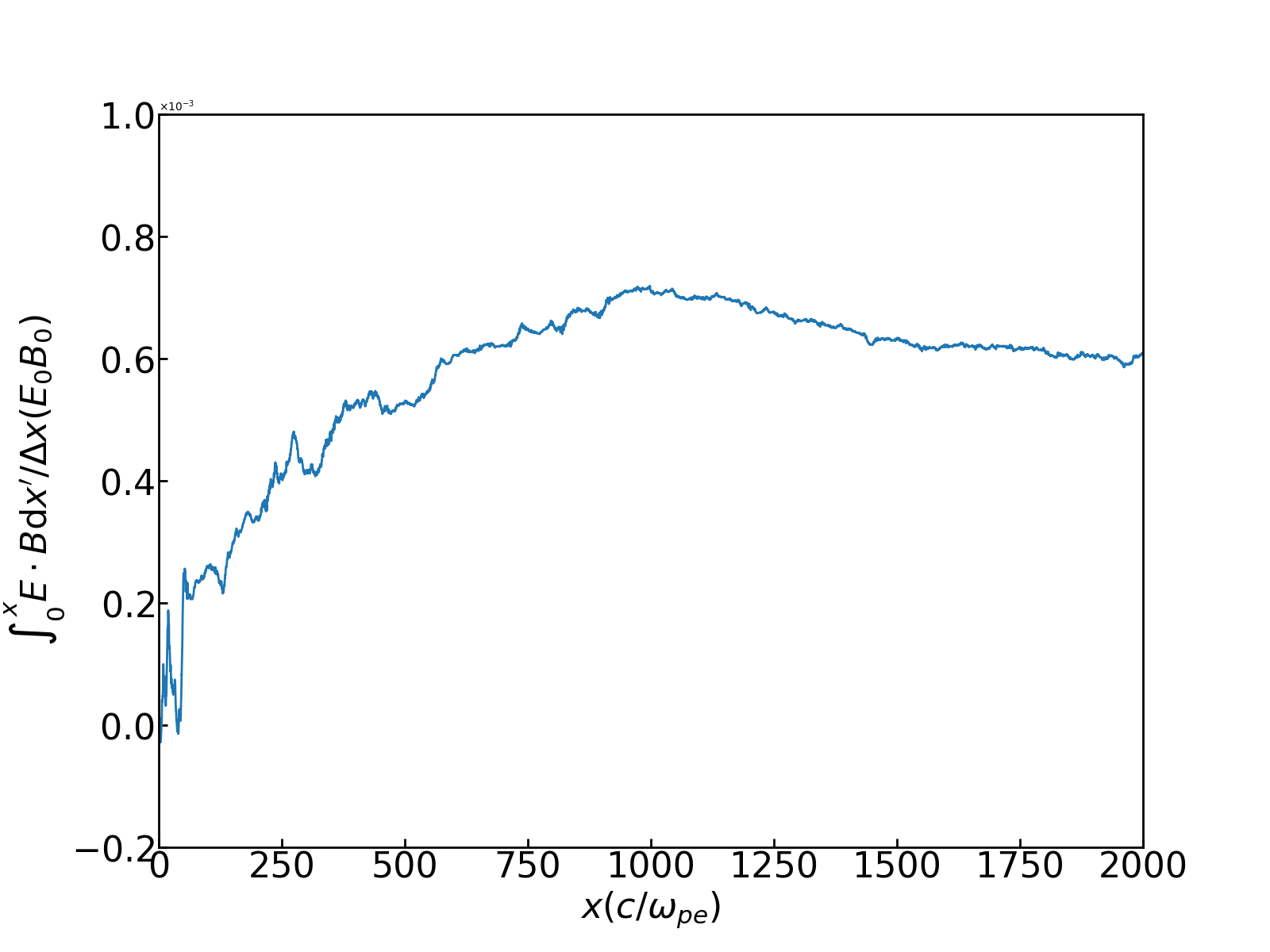}
        \caption{Distribution of $\mathbf{E}\cdot\mathbf{B}$ on $x$ (left) and the cumulative distribution of $\mathbf{E}\cdot\mathbf{B}$, i.e., $\int_{0}^{x} \mathbf{E} \cdot \mathbf{B} {\rm d}x' /\Delta x$ with $\Delta x = 2000\,c/\omega_{\mathrm{pe}}$ (right)  in the upstream region of the shock at $t=2.4\times10^4\,\omega_{\mathrm{pe}}^{-1}$. }
        \label{fig:eb}
\end{figure*}

Fig.\ref{fig:shockstruc24} depicts the spatial distributions of proton number density and self-generated transverse magnetic fields ($\delta B_{y}=B_y - B_0\sin\theta$ and $\delta B_{z}=B_z$) at $t=2.4\times10^4\,\omega_{\mathrm{pe}}^{-1}$. At this time, as the plasma approaches the left boundary $x=0$ and is reflected back into the incoming medium, a shock is generated at $x_{\mathrm{sh}}=1.5\times 10^2\,c/\omega_{\mathrm{pe}}$. In most of the downstream region, the proton number density experiences a compression by a factor of approximately $r\approx4$ compared to the value far upstream of the shock, which aligns with the anticipated outcome.
Fig.\ref{fig:phase28} indicates the particle distributions in the phase space of $x-p_{x}$.  Some particles are hindered by shock potential energy or the influence of magnetic mirrors, and are reflected back upstream of the shock near the shock front with $p_x>0$.
The reflected positrons extend over a distance greater than $8000\,c/\omega_{\mathrm{pe}}$ in the $x$ direction, which is much larger than that of electrons, i.e., $\sim4000\,c/\omega_{\mathrm{pe}}$.  Upstream of the shock, as the reflected particles stream relative to the magnetic field lines, magnetic waves are generated. These transverse magnetic fields, which can also lead to particle reflection through magnetic mirroring \citep{2009A&ARv..17..409T,2016ApJ...820...21S,2018ApJ...864..105H}, are effectively excited in the foreshock region. Consequently, the particle density upstream of the shock undergoes modulation due to the presence of these waves.

Electrically charged particles gain energy after the reflection at the shock, which is crucial for further accelerating the particles. In a nonrelativistic shock within a plasma consisting of protons and electrons, protons can undergo efficient acceleration through DSA. On the other hand, electrons require preacceleration within the shock, where their momentum is increased by several orders of magnitude, before they can undergo DSA \citep{2009A&ARv..17..409T,2015PhRvL.114h5003P,2022ApJ...931..129M}. Protons initially acquire energy through a few gyrocycles of SDA, and eventually, they are injected into the DSA regime \citep{2015PhRvL.114h5003P}. As an alternative mechanism, electrons can be preaccelerated through SDA and by scattering on Bell waves in the nearest upstream region before being injected into DSA \citep{2015PhRvL.114h5003P}. It can be seen that in the upstream regions, particles have two components, with the higher-energy part ($p \geq 0.5 m_{\mathrm{e}}c$ for electrons and positrons, $p \geq 8 m_{\mathrm{e}}c$ for protons) originated from the accelerated particles. The spatial distribution of $\mathrm{E}\cdot\mathrm{B}$ upstream of the shock and its cumulative distribution ($\int_{0}^{x} \mathbf{E} \cdot \mathbf{B} {\rm d}x' /\Delta x$ with $\Delta x = 2000\,c/\omega_{\mathrm{pe}}$) are shown in Fig.\ref{fig:eb}. Near the shock in the upstream region, the positive peak value of  $\mathrm{E}\cdot\mathrm{B}$ is greater than the negative peak one. Moreover, the places where the dot product $\mathbf{E}\cdot\mathbf{B}$ is positive outnumber the places where it is negative. This implies that when the reflected positrons move roughly parallel to the  magnetic field in the upstream of the shock, they can be accelerated by the electric field, thereby gaining energy. This acceleration process for positrons has been discovered in previous kinetic simulations on positron acceleration by shock waves in an electron-positron-ion plasma \citep{2005PhPl...12a2312H,2005PhPl...12h2306H}. As a result, the positrons reflected by the shock can obtain energy with momentum above $10 m_{\mathrm{e}}c$ at $t=2.4\times10^4\,\omega_{\mathrm{pe}}^{-1}$, and the positrons are more efficient at acquiring energy than electrons during the motion along the magnetic field upstream of shock after the reflection by the shock.

\begin{figure}
        \centering
        \includegraphics[width=0.5\textwidth]{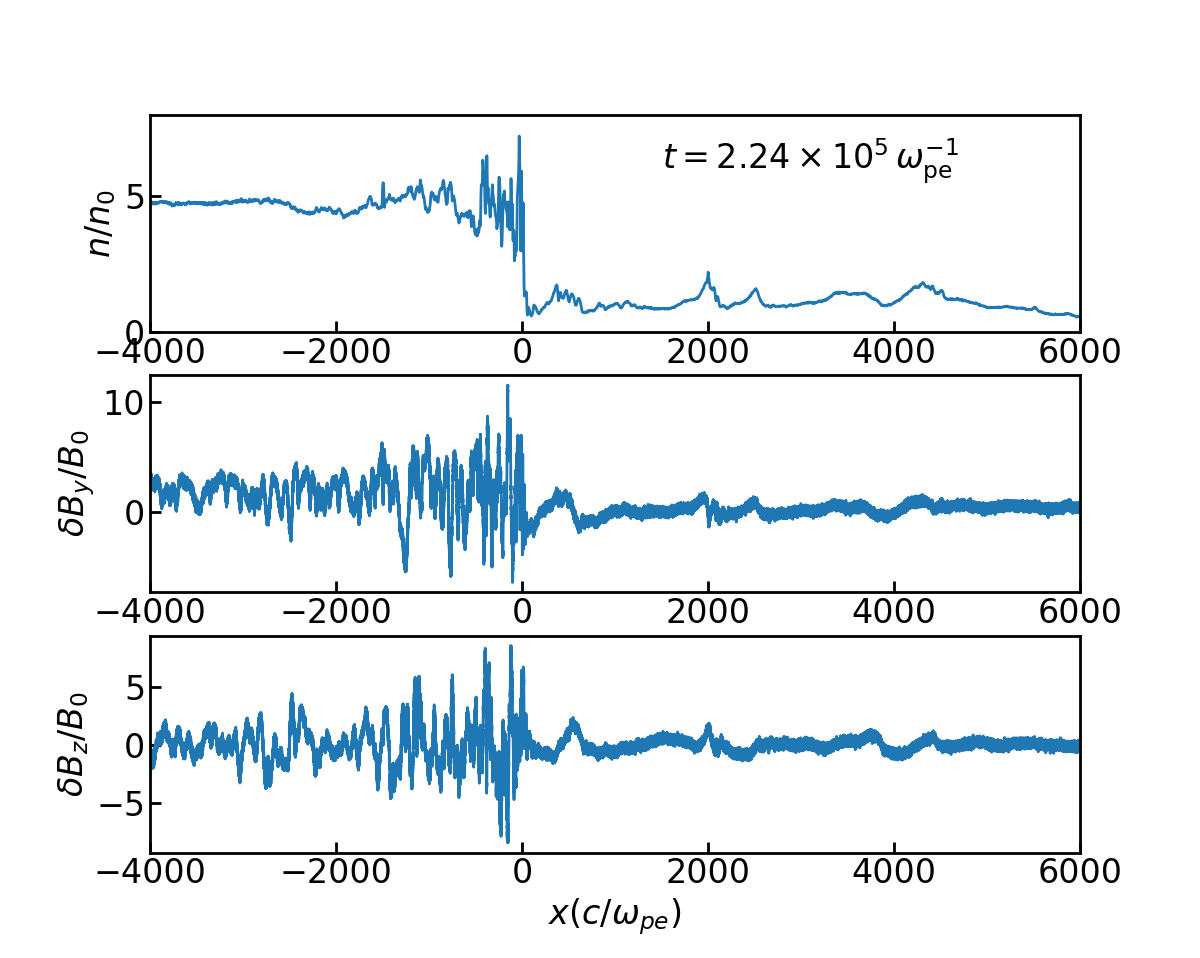}
        \caption{Proton number density normalized by $n_0$ (top panel), the self-generated magnetic field $\delta B_y$ (middle panel) and $\delta B_z$ (bottom panel) for the shock at $2.24\times10^5\,\omega_{\mathrm{pe}}^{-1}$. The shock is located at $x_{\mathrm{sh}}=1.22\times 10^4\,c/\omega_{\mathrm{pe}}$, and the $x$ coordinate is shifted by setting  the shock location as the origin. }
        \label{fig:shockstruc224}
\end{figure}

\begin{figure}
        \centering
        \includegraphics[width=0.5\textwidth]{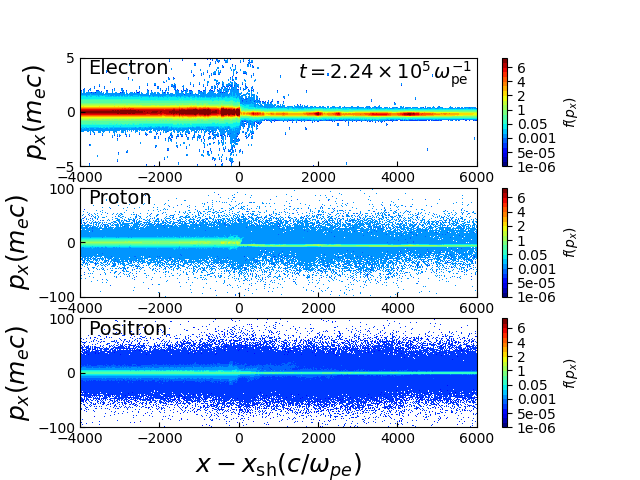}
        \caption{Phase space distributions of the electrons (top panel), the protons (middle panel), and the positrons (bottom panel) for the shock at $t=2.24\times10^5\,\omega_{\mathrm{pe}}^{-1}$.}
        \label{fig:phase224}
\end{figure}

\begin{figure*}
        \centering
        \includegraphics[width=1.0\textwidth]{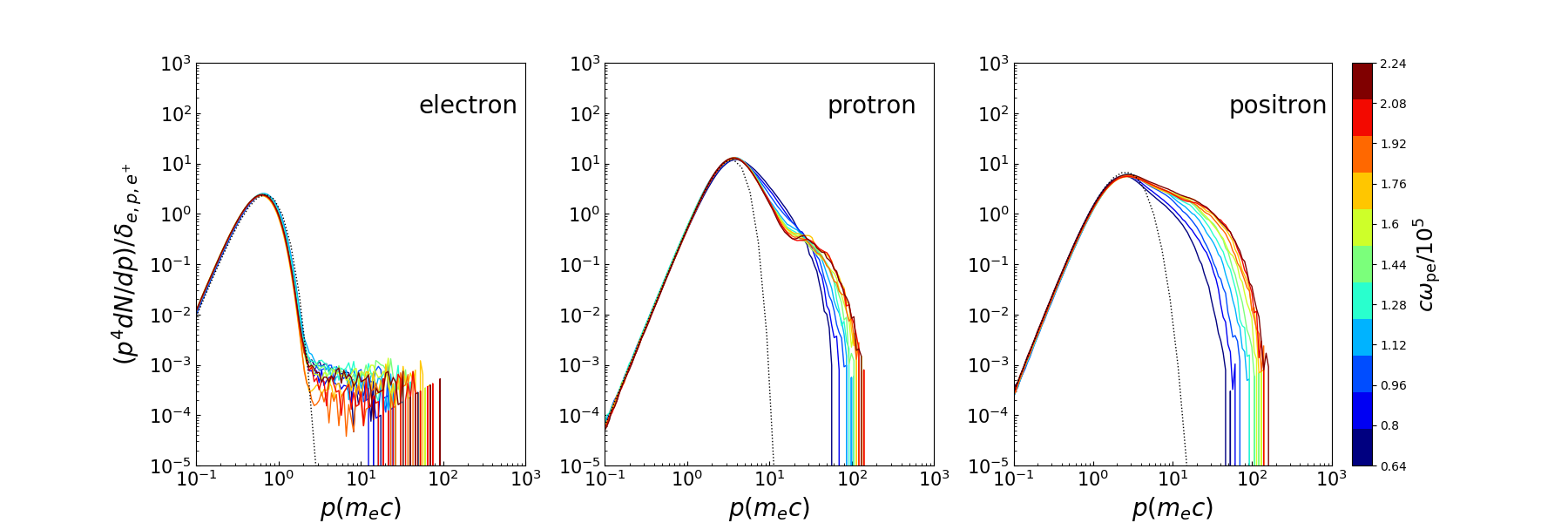}
        \includegraphics[width=1.0\textwidth]{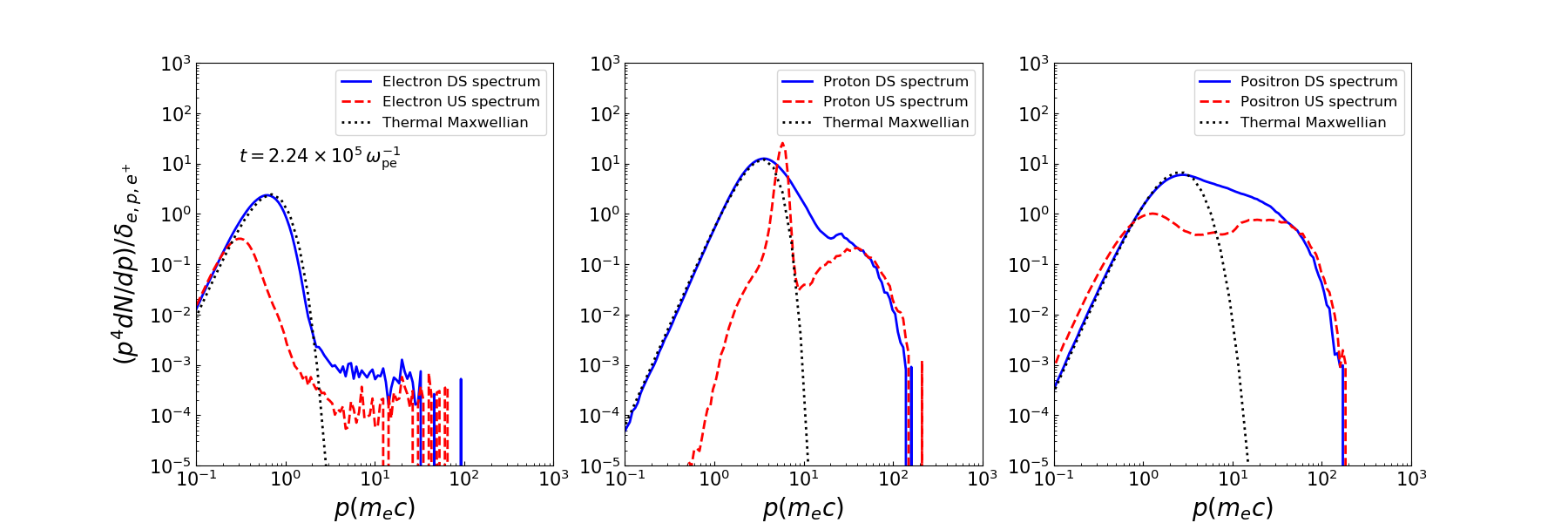}

        \caption{Upperpanels: Evolution of the downstream ($200-2200\,c/\omega_{\mathrm{pe}}$ behind the shock) normalized spectra for electrons, protons, and positrons, respectively.
         Lower panels: Normalized spectra of electrons (left), protons (middle), and positrons (right), in the region $200-2200\,c/\omega_{\mathrm{pe}}$ behind (downstream) and before (upstream) the shock at $t=2.24\times10^5\,\omega_{\mathrm{pe}}^{-1}$, respectively.  The thermal Maxwellian distributions with temperatures of $T_{\mathrm{e}}=0.1\,m_{\mathrm{e}}c^2/k_{\mathrm{B}}$, $T_{\mathrm{p}}=0.1\,m_{\mathrm{e}}c^2/k_{\mathrm{B}}$ and $T_{\mathrm{e^+}}=0.6\,m_{\mathrm{e}}c^2/k_{\mathrm{B}}$ are illustrated as  dotted lines for electrons, protons and positrons, respectively. The spectra are normalized with $\delta_e = 1.1$, $\delta_p = 1$, $\delta_{e^+} = 0.1$.}
        \label{fig:ParticleDis224}
\end{figure*}

The spatial distributions of proton number density and the self-generated transverse magnetic field at $t=2.24\times10^5\,\omega_{\mathrm{pe}}^{-1}$ are depicted in Fig.\ref{fig:shockstruc224}, with the shock located at $x_{\mathrm{sh}}=1.22\times 10^4\,c/\omega_{\mathrm{pe}}$. Efficiently generated magnetic waves both upstream and downstream of the shock cause significant scattering of charged particles in the plasma. As shown in Fig.\ref{fig:phase224}, a large number of protons and positrons are observed to point towards the $-x$ direction at this particular time, which is a result of their interaction with the self-generated magnetic waves.

The evolution of the downstream normalized spectra for each species of the particles in the region $200-2200\,c/\omega_{\mathrm{pe}}$ behind the shock and these of electrons, protons, and positrons in the region of $200-2200\,c/\omega_{\mathrm{pe}}$ behind (downstream) and before (upstream) the shock at $t=2.24\times10^5\,\omega_{\mathrm{pe}}^{-1}$ with $\delta_e = 1.1$, $\delta_p = 1$, $\delta_{e^+} = 0.1$ are presented in Figure \ref{fig:ParticleDis224}. The dotted lines represent the thermal Maxwellian distributions with temperatures of $T_{\mathrm{e}}=0.1\,m_{\mathrm{e}}c^2/k_{\mathrm{B}}$, $T_{\mathrm{p}}=0.1\,m_{\mathrm{e}}c^2/k_{\mathrm{B}}$ and $T_{\mathrm{e^+}}=0.6\,m_{\mathrm{e}}c^2/k_{\mathrm{B}}$ for electrons, protons and positrons, respectively.  The relationship between the peak momentum of the spectra  ($p^4 f(p) \propto p^4 \exp(-(m^2 c^4 + p^2 c^2)^{1/2}/k_{\mathrm{B}}T)) $ for the thermal distribution and the temperature of particles with mass of $m$ is $p_{\mathrm{k}}/mc = (8\vartheta^2+4\vartheta(1+4\vartheta^2)^{1/2})^{1/2}$, where $\vartheta=k_{\mathrm{B}}T/mc^2$.  As illustrated in Fig.\ref{fig:ParticleDis224}, the thermal distributions for the electrons peaks at  $p=0.7\,m_\mathrm{e} c$ with a temperature of $0.1\,m_{\mathrm{e}}c^2/k_{\mathrm{B}}$. The peak momentum of the positrons is equal to that of the protons, i.e., $p_{\mathrm{k}}\sim 3\,\,m_\mathrm{e}c$, which corresponds to a temperature of  $0.6\,m_\mathrm{e}c^2/k_{\mathrm{B}}$ for the positrons.  For the thermal protons with a temperature of $0.1\,m_{\mathrm{e}}c^2/k_{\mathrm{B}}$ ($\vartheta = 3.3\times10^{-3} \ll 1 $), the spectra ($p^4 f(p)$) peaks at $p_k \approx 2 \vartheta^{1/2} m_{\mathrm{p}}c = 3.5m_{\mathrm{e}}c$.  In the
downstream frame, the upstream kinetic energy density is $\sim 0.5 n_0 m_p V_0^2 = 0.3 n_0 m_{\mathrm{e}}c^2$, and the corresponding energy contained by these particles for the thermal distribution in the simulation is $\sim (\delta_e T_{\mathrm{e}} + \delta_p T_{\mathrm{p}} + \delta_{e^+} T_{\mathrm{e^+}} ) n_0 m_{\mathrm{e}}c^2 = 0.27 n_0 m_{\mathrm{e}}c^2$. The difference can mainly be attributed to the transfer of a portion of the kinetic energy to the nonthermal particles.

All three species of particles have undergone acceleration in the shock, leading to the emergence of a high-energy tail in addition to the Maxwellian distribution. With the accumulation of time, more particles are accelerated into the high-energy tail, and the maximum energy of particles gradually increases.  In the downstream region, the electron spectrum follows a power law ($dN/dp \propto p^{-s}$) with an index of $s\sim 4$ within the range of $p=3 - 10 m_{\mathrm{e}}c$. However, the spectra for protons and positrons in the high-energy tail ($p>\sim 10\,m_{\mathrm{e}}c$) exhibit relatively soft distributions with $s>4$, suggesting that at $t=2.24\times10^5\,\omega_{\mathrm{pe}}^{-1}$, the establishment of diffusive shock acceleration is incomplete due to time constraints. We define the acceleration efficiency $\eta$ based on the ratio of the energy  of given species with momentum  above ($p_{\mathrm{inj}}$) to the total energy  carried by particles of all species. At $t=2.24\times10^5\,\omega_{\mathrm{pe}}^{-1}$, the efficiencies for the electrons , protons, and positrons in the downstream region $200-2200\,c/\omega_{\mathrm{pe}}$ behind the shock are $0.012\%$ ($p_{\mathrm {inj}} = 2m_{\mathrm{e}}c$), $14.7\%$  ($p_{\mathrm {inj}} = 5m_{\mathrm{e}}c$), and $1.6\%$ ($p_{\mathrm {inj}} = 5m_{\mathrm{e}}c$), respectively. The value of $p_{\mathrm {inj}} $ for each type of particle  is determined based on the spectrum downstream of the shock. The nonthermal acceleration efficiency of positrons is about an order of magnitude lower than that of protons, mainly because the number of positrons is only $10\%$ of that of protons.
In the upstream region, there are two distinct components for each particle species: the incoming component and the accelerated one. Consequently, the spectra of protons and positrons in the upstream region exhibit a bimodal structure.

The acceleration efficiency of electrons in the shock is relatively lower compared to protons and positrons. Despite the fact that the number of the incoming positrons are only about one order of magnitude smaller than the incoming electrons, the nonthermal positrons in the downstream region ($200-2200\,c/\omega_{\mathrm{pe}}$ behind the shock)  outnumber the electrons by approximately two orders of magnitude.
In a supercritical quasi-parallel shock, the transverse magnetic field is excited in the fore shock region, causing the shock to become quasi-perpendicular. Protons first acquire energy through shock drift acceleration as they drift along the motional electric field. Once they have gained sufficient energy, they can be injected into the diffusive shock acceleration regime \citep{2015PhRvL.114h5003P,2016ApJ...820...21S}.

In the downstream region, the protons and electrons exhibit a thermal distribution with a temperature of approximately $0.1  m_{e}c^2/k_{B}$, while the positrons are heated to a temperature of about $0.6  m_{e}c^2/k_{B}$. The thermal distribution for the positrons has a peak momentum of $\sim 3 m_{e}c$, which is comparable to that of the protons. Additionally, the maximum momentum achieved by the accelerated positrons is similar to that of the protons. Fig.\ref{fig:phase224} shows that the distribution of positrons in phase space is comparable to that of the protons, indicating that the positrons undergo a similar acceleration process as the protons. A significant fraction of positrons can also undergo specular reflection at the shock and experience heating to a higher temperature as they interact with the magnetic field during their flow from upstream to downstream..

\section{Summary and discussion}
\label{sec:summary}

In this paper, we investigate the acceleration of particles consisting of protons, electrons, and positrons in a nonrelativistic quasi-parallel shock with Mach numbers of $M_{\mathrm{s}} = 17.1$ and $M_{\mathrm{A}} = 16.8$. The positron-to-electron number ratio is 0.09, which is similar to what is observed in cosmic rays at around 1 GeV. Our findings indicate that the shock structure remains largely unchanged when considering the presence of positrons. Positrons acquire higher energy due to the acceleration by the electric field parallel to the magnetic field upstream of the shock when they move along the magnetic field after the reflection by the shock, making them more easily injected into the diffusive shock acceleration process compared to electrons.
As a result, we observe that during their interaction with the magnetic field, positrons can attain higher energies and be injected more efficiently into the acceleration process compared to electrons. Consequently, we conclude that positrons achieve greater acceleration efficiency than electrons within the shock. This discovery holds significant implications for comprehending the origin and composition of cosmic rays, as well as their interactions with the interstellar medium..

As indicated by the observations from PAMELA \citep{2009Natur.458..607A}, Fermi-LAT \citep{2012PhRvL.108a1103A}, and AMS-02 \citep{2013PhRvL.110n1102A}, there is an excess of positron flux in cosmic rays with energies greater than or equal to 10 GeV. The origin of this excess remains uncertain, and several possible explanations have been proposed, including nearby pulsars \citep{2022CoTPh..74j5403Z}, pulsar wind nebulae (PWNe) \citep{2023PhRvD.107l3020S}, and supernova remnants (SNRs) \citep{2013APh....49...23E}. Our simulation in this paper suggests that the acceleration of positrons in quasi-parallel nonrelativistic shocks is efficient, since they can be injected into the DSA process more effectively than electrons. It is supported the hypothesis that SNRs may serve as sources of positrons in cosmic rays within the Galaxy.

The simulations of particle acceleration at quasi-parallel nonrelativistic shocks are rather  time-consuming, and we adopted a reduced proton-to-electron mass ratio of $m_{\mathrm p}/m_{e} = 30$ in this paper to obtain the acceleration characteristics of particles earlier in the numerical simulation process. With a larger proton-to-electron mass ratio, a wider simulation domain and a longer computation time are required. In the shocks within the plasma composed primarily of electrons and protons, electrons must be pre-accelerated to obtain a large enough Larmor radii comparable to the shock width before being injected into the DSA. Alternatively, a fraction of positrons that flow towards the shock front are reflected by the shock, and the reflected positrons can be further accelerated by the electric field parallel to the magnetic field upstream of the shock, resulting in a significant increase in their momentum. This makes them more susceptible to injection into the DSA process than electrons, and the effect is not influenced by the reduced proton-to-electron mass ratio. In this paper, we investigated the acceleration process of electrons, protons, and positrons in a specific shock wave in which the sonic and Alfv\'{e}n Mach numbers almost coincide, but this coincidence is not necessary. The number of particles accelerated by shock waves can be influenced by the proton-to-electron mass ratio, the shock Mach numbers, and the shock obliquity. However, these factors are not investigated in this paper.

\section*{acknowledgements}

This research is supported by NSFC through Grants 12063004 and 12393852, as well as grants from the Yunnan Provincial Government (YNWR-QNBJ-2018-049) and Yunnan Fundamental Research Projects (grant No. 202201BF070001-020).

\bibliographystyle{aasjournal}
\bibliography{posiacc}
\end{document}